# A High Efficiency Ultra High Vacuum Compatible

# Flat Field Spectrometer for EUV Wavelengths

B. Blagojevic[1], E.-O. Le Bigot[2], K. Fahy[3], A. Aguilar[1], K. Makonyi[1], E. Takács[1],

J.N. Tan[1], J.M. Pomeroy[1], J.H. Burnett[1], J.D. Gillaspy[1] and J.R. Roberts[1]

[1]Atomic Physics Division, National Institute of Standards and Technology (NIST),

Gaithersburg, MD 20899, USA

[2]Laboratoire Kastler Brossel, Université P. & M. Curie et Ecole Normale Supérieure,

Case 74, 4 place Jussieu, 75005 Paris, France

[3]Department of Experimental Physics, University College Dublin, Belfield, Dublin 4,

Ireland

ABSTRACT

A custom, flat field, extreme ultraviolet EUV spectrometer built specifically for use with low power light sources that operate under ultrahigh vacuum conditions is reported. The spectral range of the spectrometer extends from 4 nm to 40 nm. The instrument optimizes the light gathering power and signal to noise ratio while achieving good resolution. A detailed description of the spectrometer and design considerations are presented, as well as a novel procedure that could be used to obtain a synthetic wavelength calibration with the aid of only a single known spectral feature. This synthetic wavelength calibration is compared to a standard wavelength calibration obtained from previously reported spectral lines of Xe, Ar and Ne ions recorded with this spectrometer.

**I. INTRODUCTION**

Interest in extreme ultraviolet (EUV) ion spectroscopy has increased in the last few years [1-15] due in part to the need of the EUV lithography community to accurately model commercial EUV light sources. EUV lithography (EUVL) is targeted by the semiconductor industry to be used as the "next generation" technology that will deliver feature sizes projected by Moore's Law. However, many critical issues need to be addressed before successful deployment can be realized. Among these is the development of powerful EUV sources that can deliver over 100 Watts in-band (+/- ≈ 0.27 nm) at 13.5 nm. In order to meet the industrial power requirement, accurate plasma models are used in support of the engineering process, which in turn require reliable fundamental EUV atomic data. As part of the effort to deliver the fundamental data needed and to provide benchmark test spectra, a custom flat-field EUV spectrometer optimized for use with the National Institute of Standards and Technology (NIST) Electron Beam Ion Trap (EBIT) has been built for EUV spectroscopy in the 4 nm to 40 nm wavelength range. While the instrument has been optimized for use with the NIST EBIT, the design is expected to provide good performance when used with almost any EUV source.

The NIST EBIT is a versatile light source, capable of producing nearly any ion charge state. The narrow electron beam energy spread allows a precise control of the charge state distribution present in the trap, and of the excitation of the trapped ions. A detailed description of the NIST EBIT may be found elsewhere [16-17]. One of the challenges of using the EBIT to support EUVL data needs is the EBIT's weak emission of EUV light

($10^{-10}$ W), compared with the EUVL sources that currently produce tens of watts. Therefore, ultra low noise and high light gathering capability were primary goals in the design of the EUV spectrometer, while also maintaining good resolution (±0.02 nm).

In designing an instrument to meet these objectives, a number of schemes for coupling the EBIT light source into the spectrometer have been extensively modeled. In section II of this paper, a detailed description of the EUV spectrometer is presented. This is followed by the calculation of the spectrometer detection efficiency in section III. The considerations in the design of this instrument are presented in section IV, in particular, the use of a spherical focusing mirror for efficient light gathering. Section V describes a synthetic calibration method based on ray-tracing calculations and a single known spectral line. Section VI presents a standard wavelength calibration based on recorded spectra of Ne, Ar and Xe ions that is compared in section VII to the synthetic calibration.

## II. APPARATUS

The EUV spectrometer is pictured in Figure 1 along with a mechanical schematic. Starting from the source end (top of figure), it consists of a zirconium vacuum window, a spherical focusing mirror, a set of bilateral continuously adjustable slits, a gold-coated-concave reflection grating, and a detector. The center of the spherical mirror is 48 cm away from the EBIT axis, at a grazing incidence angle of 3°. The spherical mirror reflecting surface is coated with gold and has a radius of curvature of 917.1 cm ± 2.3 cm. The physical dimensions of the mirror are 4 cm high by 10 cm long. The function of this mirror is to collect light from the EBIT source and concentrate it onto the bilateral, continuously adjustable slits, located 48 cm from the center of this mirror. The reflection grating follows the 1200 lines/mm design reported by Harada and Kita [18] and has been further characterized by several researchers [19,20]. The grating has a radius of curvature of 564.9 cm ± 2.0 cm, variable groove spacing (a flat field grating) and is 3 cm high by 5 cm long. This grating has been implemented in other similar EUV spectrometers [2,15,21,22]. In the NIST spectrometer, the grating is located 23.7 cm from the slits and is placed at a grazing incidence angle of 3°. A mask is placed in front of the grating that reduces the amount of scattered light impinging on the grating. The detector is a liquid nitrogen cooled, back illuminated charge coupled device (CCD) camera that is placed in the focal plane of the reflection grating, located 23.5 cm away from the grating center. The dimensions between the slits, grating, and detector are the same as those used by Kita et al. [23], yielding a reciprocal linear dispersion of the grating that varies from 4.24 nm/cm at 5 nm to 7.55 nm/cm at 20 nm in the first negative diffractive order. The CCD array consists of 1340 x 400 pixels (each pixel is 20 $\mu$m x 20 $\mu$m) and is directly exposed to the EUV radiation. It is mounted on a two dimensional linear stage. One dimension allows the surface of the CCD array to be positioned in the focal plane of the grating, by adjusting the distance from the grating, while the other dimension is used to select the wavelength range. The angular position of the CCD plane is fixed with respect to the vacuum chamber and the diffraction grating is adjusted (with a retro-reflecting laser beam) such that the focal plane is parallel to the CCD face within 3 arc minutes. Ultra-high vacuum is maintained by two 100 L/s ion pumps connected to the mirror chamber and the grating chamber. The pressure in each chamber is in the $10^{-7}$ Pa range during data acquisition, as measured by cold cathode gauges. Cooling the CCD camera with liquid nitrogen also helps reduce the gas load in the grating chamber. Neither UHV chamber was baked after the installation of fragile optics. A 0.1 μm thick zirconium foil, supported by a 70 lines/mm nickel mesh with 87 % open area, is mounted in a valve on the optical axis between the EBIT light source and the focusing mirror. This foil serves as a filter to block visible light, and also provides the option of isolating the vacuum of the EBIT (typically $\approx 10^{-8}$ Pa) from that of the spectrometer. The Zr filter efficiently transmits EUV radiation in the band between 5 nm and 25 nm (discussed further in the next section) [24].

## III. SPECTROMETER DETECTION EFFICIENCY AND RESOLVING POWER.

The detection efficiency of the spectrometer is defined as the ratio of the number of photons detected by the CCD array to the number of photons entering the mirror's numerical aperture. The detection efficiency was calculated in negative first diffractive order using the ray tracing code SHADOW [25]. In this calculation, the dimensions listed in Table I and the efficiency of each component were included, except the quantum efficiency of the CCD array and the transmission of the Zr-filter, which were incorporated separately. The detection efficiency as a function of wavelength is shown in Figure 2, as determined by launching rays uniformly over the range of source positions and angles that accurately model the EBIT with a 300 μm slit width; the dashed line in Figure 2 shows the effect of the Zr filter on the detection efficiency. While the relative efficiency as calculated by SHADOW is expected to be representative of the instrument, SHADOW overestimated the absolute efficiency of the grating (SHADOW predicts about 80%) when compared with experimental measurements, which report absolute efficiencies of about 10% in this wavelength range [26,27]. The second most significant factor in the instrument's detection efficiency is the quantum efficiency of the CCD camera, which varies between 40 % and 50 % in the 4 nm to 40 nm range. A resolving power ($\lambda/\Delta\lambda$, where $\lambda$ is the wavelength and $\Delta\lambda$ the resolution) of 577 for 50 $\mu$m slit size was obtained by measuring the $4d^{10}\ ^1S_0$-$4d^9 4f\ ^1P_1$ Xe IX line profile. For slit sizes larger than 300 $\mu$m the resolving power remains at a constant value of 350.

## IV. DESIGN CONSIDERATIONS

In order to maximize the light gathering power of the spectrometer, many systems were studied and simulated using the ray-tracing code SHADOW [25] through the SHADOW Visual User Interface [28] (both programs are freely available). Calculations of the reflectivity of gold-coated mirrors at 4 nm show a drop with increasing angle to near zero at a grazing incidence angle of about 10 degrees [24]. This limits the selection of light gathering systems to those at grazing incidence angles of < 10º. The throughput was maximized by optimizing the intensity of the light reaching the simulated CCD array. For each of the light collection systems considered, the distance between the EBIT source and the (first) mirror was kept fixed (to the smallest possible distance), while many possible incidence angles and mirror radii of curvature were tried.

Among the systems considered was the Kirkpatrick-Baez two-spherical-mirror scheme [29] previously employed in a flat field spectrometer [22] used for laser produced plasma studies. In the case of a weak light source (e.g., EBIT source), this system presents the disadvantage of having intrinsic loss of light collection due to the small solid angle of the first mirror as seen from the second mirror. Other systems based on a single mirror of various geometries were also considered, such as toroidal, ellipsoidal, and elliptical cylinder geometries. The toroidal mirror has very poor sagittal focusing unless the smaller of the two radii of curvature is made very small compared to the distance between source and mirror. In this limit, substantial aberrations are introduced. The ellipsoidal and the elliptical cylinder mirror would result in very efficient light collection systems if the mirror is placed at about 10 cm from the EBIT axis. However, this would require positioning the mirror in a high-voltage environment representing substantial practical difficulties.

The optimal light gathering system was found to be a single spherical mirror placed at 48 cm from the EBIT axis at a grazing incidence angle of 3°. This configuration maximizes the solid angle as seen from the EBIT while ensuring high reflectivity and preventing overfilling of the grating. This system concentrates the light emitted from the EBIT onto the slits, although the design also works well for a wide variety of other types of sources. In all these simulations the source dimensions were 2.5 cm x 300 μm x 300 μm, which are typical of the EBIT's plasma dimensions.

## V. SYNTHETIC CALIBRATION

For many applications other than the EBIT, easy identification of spectral markers is not possible due to the abundance (or scarcity for synchrotron sources) of spectral features.

The calibration procedure presented here provides a method for determining spectral positions accurately from a single known spectral line.

The complete system is modeled in SHADOW for each spectral feature by launching tens of thousands of rays at a distribution of starting positions and angles mimicking the geometry of the source. The rays are distributed in the simulated exit plane of the spectrometer over a width that depends on the slit width. The center position of the distribution is selected by binning the exit plane positions and assigning the position of the maximum bin $x_i$ to correspond to the input wavelength $\lambda_i$. Spectral lines were simulated from 4 nm to 40 nm in 1 nm increments to form a set of points $(x_i, \lambda_i)$, which are plotted as solid dots in Figure 3. The x=0 position is at the intersection of the line that runs from the center of the grating perpendicular to the focal plane, shown in Figure 4. The dimensions and critical quantities for accurate modeling of the spectrometer using SHADOW are included in Table I. Since a simple, theoretical dispersion function for the flat-field grating is not known, a fourth order polynomial expansion was used to fit the set of points $(x_i, \lambda_i)$:

$$\lambda_{syn}(x) = \sum_{i=0}^{4} a_i x^i \; , \tag{1}$$

where $\lambda_{syn}$ is the wavelength in nm, $x$ is the distance along the grating focal plane in cm and $a_0 = -1.179$ nm, $a_1 = 0.01753$ nm/cm, $a_2 = 0.7605$ nm/cm$^2$, $a_3 = -0.00317$ nm/cm$^3$, $a_4 = -0.00064$ nm/cm$^4$ are the coefficients for the function plotted as a solid line in Figure 3.

The location of the CCD on the focal plane is found by recording the pixel number $p_k$ of one known spectral line $\lambda_k$ as shown in Figure 4. The corresponding position $x_k$ of this pixel is found by use of SHADOW. In our case, the CCD camera used has a pixel size of 20 μm, i.e. 500 pixels/cm. Thus, the position in centimeters $x_0$ of the first pixel is $x_0 = x_k - p_k/500$. The position of each pixel $x(p)$ on the focal plane is

$$x(p) = x_0 + \frac{p}{500} \qquad , \qquad (2)$$

where $x(p)$ and $x_0$ are in centimeters. The wavelength as function of pixel number $\lambda_{syn}(p)$ is:

$$\lambda_{syn}(p) = \sum_{i=0}^{4} a_i \left( x_0 + \frac{p}{500} \right)^i \qquad . \qquad (3)$$

As an example of the above procedure, a synthetic calibration was obtained for the EUV spectrometer by using the $4d^{10}\ ^1S_0 - 4d^9 4f\ ^1P_1$ transition of Xe IX as the known line. This line has been reported by Churilov and Joshi [30] at $\lambda_k = 12.0133$ nm ± 0.003 nm. This transition is the most intense line observed with the EBIT in the 4 nm to 20 nm spectral range. Then, the corresponding $p_k$ was found by a peak fitting procedure to be at 773.7 pixels and the calculated position $x_k$ at 4.223 cm.

## VI. STANDARD CALIBRATION

Spectra of Xe, Ar and Ne trapped ions at the EBIT were recorded in the 4 nm to 20 nm range at 8 keV electron beam energy. These measurements are shown in Figures 5-7. The numbers on the figures correspond to 35 previously reported lines listed in Table II and used to perform a standard calibration. For this calibration, each spectral feature indicated in Figures 5-7 was fitted with a Voigt function to determine the pixel center (non-integer pixel number). These values were plotted versus their corresponding reported wavelengths and fitted with a fourth order polynomial to obtain the standard wavelength calibration ($\lambda_{std}$). The coefficients of the polynomial are $b_0$=4.22371 nm, $b_1$=0.00799 nm, $b_2$=2.7413x10$^{-6}$ nm, $b_3$=-7.42166x10$^{-11}$ nm, and $b_4$=-1.11772x10$^{-14}$ nm. The columns in Table II are, respectively, the feature number, identified transition, reported wavelengths ($\lambda_{rep}$), bibliography reference, pixel center, the wavelength obtained from the synthetic calibration ($\lambda_{syn}$) given by equation 3 and the last column contains the standard calibration wavelengths ($\lambda_{std}$).

All recorded spectra used for the wavelength calibration were acquired using the CCD camera operating in spectroscopy mode. In this mode, the CCD array of 1340 x 400 pixels (horizontal x vertical) is converted to a one-dimensional row of 1340 pixels by hardware binning along the vertical dimension (sagittal) using the manufacturer's acquisition software. The spectra recorded in this way has an improved signal to noise ratio (S/N) for a given integration time due to the factor of 400 reduction in readout noise, compared to the spectra acquired with no binning and analyzed as a full 1340 x 400 array. The cosmic radiation background was removed by using multiple frames in post-acquisition data processing. Prior to the wavelength calibration measurements, ion spectral lines were acquired by the CCD array operating in imaging mode (no binning) to establish that all recorded lines were as parallel as possible to the vertical 400-pixel rows and did not exhibit any observable optical aberrations (e.g., curvature).

## VII. DISCUSSION.

Shown in Figure 8 are the differences between the synthetic and standard wavelengths from the reported values ($\lambda_{syn}$-$\lambda_{rep}$ and $\lambda_{std}$-$\lambda_{rep}$) for the 35 identified lines [30-39]. The standard deviations of the differences are 0.011 nm and 0.008 nm respectively. The gray region in Figure 8 represents the spectral width of a single pixel (+/- _ pixel) that varies from 0.008 nm to 0.015 nm at 4 nm and 20 nm respectively. The error bars in the plot are the uncertainties reported in the literature for each $\lambda_{rep}$ and represent the uncertainty at each point for both series, shown for just one series to reduce clutter in the plot. The reduced accuracy of the synthetic calibration is due in part to the difficulty of simulating factors such as fabrication tolerances, the difficulty in positioning the CCD in the theoretical focal plane of the grating and the final alignment of the optical elements (including CCD rotational angle). The discrepancy between the synthetic and the standard calibration can be accounted for with a small linear correction to the synthetic calibration. The use of the synthetic calibration allows users to deploy the instrument expediently with minimal loss of accuracy. For maximum precision the user is encouraged to perform a standard calibration.

In this paper, a flat field, UHV extreme ultraviolet spectrometer with good resolution and optimized light collection has been presented along with a demonstration of performance. Further, a novel technique for calibrating this (and potentially other spectrometers) based on a single known spectral feature and ray tracing is presented.  Comparison with the standard calibration procedure indicates a total uncertainty similar to the size of the CCD pixel, probably due to the peak selection technique used in the synthetic calibration.

**Acknowledgments**

We thank Jorge Rocca, Glenn Kubiak, Martin Richardson, Greg Shimkaveg, Wayne McKinney, Enrique Parra, Howard Milchberg, Steven Grantham, and Charles Tarrio for helpful advice.  This work was partially supported by International SEMATECH under LITH152.

# REFERENCES

1. M. McGeoch, Appl. Opt. **37**, 1651 (1998).

2. M. A. Klosner, and W. T. Silfvast, Opt. Lett. **23**, 1609 (1998).

3. K. Bergmann, G. Schriever, O. Rosier, M. Müller, W. Neff and R. Lebert, Appl. Opt. **38**, 5413 (1999).

4. M. A. Klosner, and W. T. Silfvast, J. Opt. Soc. Am. B **17**, 1279 (2000).

5. K. Bergmann, O. Rosier, W. Neff and R. Lebert, Appl. Opt. **39**, 3833 (2000).

6. I. Krisch, P. Choi, J. Larour, M. Favre, J. Rous and C. Leblanc, Contrib. Plasma Phys. **40**, 135 (2000).

7. T. Boboc, R. Bischoff and H. Langoff, J. Phys. D **34**, 2512 (2001).

8. E. Robert, B. Blagojevi_, R. Dussart, S. R. Mohanty, M. M. Idrissi, D. Hong, R. Viladrosa, J.-M. Pouvesle, C. Fleurier and C. Cachoncinlle, Proc. SPIE **4343**, 566 (2001).

9. C. Biedermann, R. Radtke, J.-L. Schwob, P. Mandelbaum, R. Doron, T. Fuchs, and G. Fußmann, Physica Scripta **T92**, 85 (2001).

10. E. Trabert, P. Beiersdorfer, and H. Chen, Phys. Rev. A, 70, 032506 (2004).

11. L. Juschkin, A. Chuvatin, S. V. Zakharov, S. Ellwi, and H.-J. Kunze, J. Phys. D **35**, 219 (2002).

12. N. R.Fornaciari, H. Bender, D. Buschenauer, J. Dimkoff, M. Kanouff, S. Karim, C. Romeo, G. Shimkaveg, W.T.Silfvast and K. D. Stewart, Proc. SPIE **4688**, 110 (2002).

13. V. M. Borisov, I. Ahmad, S. Göetze, A. S. Ivanov, O. B. Khristoforov, J. Kleinschmidt, V. Korobotchko, J. Ringling, G. Schriever, U. Stamm and A. Y. Vinokhodov, Proc. SPIE **4688**, 626 (2002).

14. N. Böwering, and M. Martins, W. N. Partlo and I. V. Fomenkov, J. Appl. Phys. **95**, 16 (2004).

15. B. M. Luther, Y. Wang, M. A. Larotonda, D. Alessi, M. Berrill, M. C. Marconi, J. J. Rocca and V. N. Shlyaptsev, *Opt. Lett.* **30**, 165 (2005).

16. J. D. Gillaspy, Y. Aglitskiy, E. W. Bell, C. M. Brown, C. T. Chantler, R. D. Deslattes, U. Feldman, L. T. Hudson, J. M. Laming, E. S. Meyer, C. A. Morgan, A. I. Pikin, J. R. Roberts, L. P. Ratliff, F. G. Serpa, J. Sugar, and E. Takács, Physica Scripta **T59**, 392 (1995).

17. J. D. Gillaspy, Physica Scripta **T7**1, 99 (1997).

18. T. Harada and T. Kita, Appl. Opt. **19**, 3987 (1980).

19. W. Schwanda, K. Eidmana and M.C. Richardson, J. X-Ray Sci. Technol., **4**, 8 (1993).

20. N. Nekano, H. Kuroda, T. Kita and T. Harada, Appl. Opt. **23**, 2386 (1984).

21. P. Beiersdorfer, J.R. Crespo Lopez-Urrutia, P. Springer, S.B. Utter and K. Wong, Rev. Sc. Inst., **70**, 276 (1999).

22. A. Saemann and K. Eidmann, Rev. Sc. Inst. **69**, 1949 (1998).

23. T. Kita, T. Harada, N. Nakano, H. Kuroda, Appl. Opt. **22**, 512 (1983).

24. E. M. Gullikson, Center for X-ray Optics web site, http://www-cxro.lbl.gov/optical_constants/

25. J.G. Chen, C. Welnak and F. Cerrina, Nucl. Instrum. and Meth. A**347**, 344 (1994), code freely available at http://www.nanotech.wisc.edu/shadow/shadow.html

26. J. Edelstein, M.C. Hettrick, S. Mrowka, P. Jelinsky, and C. Martin, *Appl. Opt.* **23***, 3267 (1984).

27. L. Poletto, G. Naletto, and G. Tondello, *Opt. Eng.* **40**, 178 (2001).

28. M. Sanchez del Rio and R. J. Dejus, SPIE proceedings, vol. **3152**, 148 (1997).

29. P. Kirkpatrick and A. V. Baez, J. Opt. Soc. Am. **38**, 766 (1948).

30. S. S. Churilov, and Y. N. Joshi, Physica Scripta **65**, 40 (2002).

31. L. W. Phillips, and W. L. Parker, Phys. Rev. **60**, 301 (1941).

32. E. Träbert, P. Beiersdorfer, J. K. Lepson, and H. Chen, Phys. Rev. A **68**, 042501 (2003).

33. J. F. Seely, C. M. Brown, U. Feldman, J. O. Ekberg, C. J. Keane, B. J. MacGowan, D. R. Kania and W. E. Behring, At. Data Nuc. Data Tables **47**, 1 (1991).

34. G. Tondello, and T. M. Paget, J. Phys. B: Atom.Molec.Phys. **3**, 1757 (1970).

35. F. W. Paul, and H. D. Polster, Phys. Rev. **59**, 424 (1941).

36. J.F. Seely, J.O. Ekberg, U. Feldman, J.L. Schwob, S. Suckewer and A. Wouters, *J. Opt. Soc. Am. B* **5**, 602 (1988).

37. S. S. Churilov, Y. N. Joshi, J. Reader and R. R. Kildiyarova, Physica Scripta **70**, 126 (2004).

38. W. A. Deutschman, and L. L. House, Astrophys. J. **144**, 435 (1966).

39. V. Kaufman, J. Sugar, W. L. Rowan, J. Opt. Soc. Am. B **5**, 1273 (1988).

**Table I.** Key spectrometer dimensions and relevant parameters needed for modeling.

| Parameters | Value |
|---|---|
| EBIT trap diameter; height | 300 μm; 2.5 cm |
| Mirror radius of curvature; height x length | 917.1 cm; 4 cm x 10 cm |
| Entrance slit width x height | 300 μm x 2.5 cm |
| Grating radius of curvature; height x length | 564.9 cm; 3 cm x 5 cm |
| Line density in the grating center (diffraction order) | 1200 lines/mm  (-1) |
| Flat field grating -- polynomial line density coefficients:  linear, quadratic, cubic | 849.71 lines/cm$^2$, 51.42019 lines/cm$^3$, 3.152668 lines/cm$^4$ |
| Distance from EBIT Center to Mirror Center | 48 cm |
| Distance from Mirror Center to Slit | 48 cm |
| Distance from Slit to Grating Center | 23.7 cm |
| Distance from Grating Center to CCD Plane | 23.5 cm |
| CCD width x height | 2.68 cm x 0.8 cm |

**Table II.** Observed transitions in Xe, Ar and Ne trapped ions, used for the wavelength calibration.

| No | Transition | $\lambda_{rep}$ (nm) | Ref. | Pixel | $\lambda_{syn}$ (nm) | $\lambda_{std}$ (nm) |
|----|------------|---------------------|------|-------|---------------------|---------------------|
| 1 | Ar IX $2p^6\ ^1S_0-2p^5 3s\ ^1P^o_1$ | 4.8730±0.002 | 31 | 78.9 | 4.861 | 4.873 |
| 2 | Ar IX $2p^6\ ^1S_0-2p^5 3s\ ^3P^o_1$ | 4.9180±0.002 | 31 | 84.4 | 4.908 | 4.918 |
| 3 | Xe XLIII $3s^2\ ^1S_0-3s3p\ ^3P_1$ | 6.2875± 0.012 | 32 | 239.5 | 6.279 | 6.288 |
| 4 | Xe XLIV $3s^2\ ^1S_0-3p^2\ ^3P_{3/2}$ | 6.658±0.03 | 33 | 279.0 | 6.650 | 6.658 |
| 5 | Ne VII $2s^2\ ^1S_0-2s4p\ ^1P^o_1$ | 7.5765±0.01 | 34 | 372.5 | 7.562 | 7.577 |
| 6 | Ne VII $2s2p\ ^3P^o_2-2s4d\ ^3D_3$ | 8.2268±0.005 | 34 | 436.0 | 8.209 | 8.227 |
| 7 | Ne VIII $2s\ ^2S_{1/2}-3p\ ^2P^o_{3/2}$ | 8.8092±0.03 | 34 | 492.1 | 8.798 | 8.809 |
| 8 | Xe IX $4d^{10}\ ^1S_0-4d^9 7p\ ^3P_1$ | 8.8444±0.007 | 30 | 495.5 | 8.835 | 8.844 |
| 9 | Ne VII $2s2p\ ^1P^o_1-2s4d\ ^1D_2$ | 8.9368±0.005 | 34 | 504.7 | 8.933 | 8.937 |
| 10 | Xe IX $4d^{10}\ ^1S_0-4d^9 5f\ ^1P_1$ | 9.6449±0.003 | 30 | 567.6 | 9.619 | 9.645 |
| 11 | Ne VII $2s^2\ ^1S_0-2s3p\ ^1P^o_1$ | 9.7502±0.005 | 34 | 578.6 | 9.741 | 9.750 |
| 12 | Ne VIII $2p\ ^2P^o_{3/2}-3d\ ^2D_{5/2}$ | 9.8260±0.005 | 34 | 585.0 | 9.812 | 9.826 |
| 13 | Ne VIII $2p\ ^2P^o_{1/2}-3s\ ^2S_{1/2}$ | 10.2911±0.005 | 34 | 627.9 | 10.296 | 10.291 |
| 14 | Ne VII $2s2p\ ^3P^o_2-2s3d\ ^3D_3$ | 10.6192±0.005 | 34 | 655.3 | 10.610 | 10.619 |
| 15 | Xe X $4p^6 4d^9\ ^2D_{5/2}-4p^6 4d^9 4f\ (870470)_{7/2}$ | 11.4880±0.003 | 30 | 729.9 | 11.485 | 11.488 |
| 16 | Ne VII $2s2p\ ^1P^o_1-2s3d\ ^1D_2$ | 11.6693±0.005 | 34 | 744.8 | 11.663 | 11.669 |
| 17 | Xe IX $4d^{10}\ ^1S_0-4d^9 4f\ ^1P_1$ | 12.0133±0.003 | 30 | 773.7 | 12.012 | 12.013 |
| 18 | Ne V $2p\ ^2P^o_{1/2}-3d\ ^2D_{3/2}$ | 12.2490±0.01* | 35 | 794.6 | 12.267 | 12.249 |
| 19 | Ne V $2p^2\ ^1D_2-2p4d\ ^1F^o_3$ | 12.2520±0.01* | 35 | 793.6 | 12.255 | 12.252 |
| 20 | Ne VII $2s2p\ ^3P^o_1-2s3s\ ^1S_0$ | 12.7663±0.01 | 34 | 834.9 | 12.765 | 12.766 |
| 21 | Xe XLIII $3s^2\ ^1S_0-3s3p\ ^3P_1$ | 12.993±0.03 | 36 | 853.4 | 12.997 | 12.993 |
| 22 | Xe XI $4d^8\ ^3P_1-4d^7 5p\ (786580)_2$ | 13.2983±0.003 | 37 | 879.3 | 13.324 | 13.298 |
| 23 | Ne V $2p\ ^2P^o_{3/2}-3s\ ^4S_{1/2}$ | 13.8640±0.01* | 35 | 921.1 | 13.860 | 13.864 |
| 24 | Ne V $2p^2\ ^3P_2-2p3d\ ^3P^o_2$ | 14.2720±0.01* | 35 | 952.3 | 14.265 | 14.272 |
| 25 | Ne V $2p^2\ ^3P_2-2p3d\ ^3D^o_3$ | 14.3305±0.01* | 35 | 957.3 | 14.330 | 14.331 |
| 26 | Ne V $2p^2\ ^1D_2-2p3d\ ^1F^o_3$ | 14.7130±0.01* | 35 | 986.4 | 14.714 | 14.713 |
| 27 | Xe X $4p^6 4d^9\ ^2D_{3/2}-4p^6 4d^9 5p\ (682998)_3$ | 15.0089±0.003 | 30 | 1009.4 | 15.020 | 15.009 |
| 28 | Xe X $4p^6 4d^9\ ^2D_{5/2}-4p^6 4d^9 5p\ (646494)_{5/2}$ | 15.4680±0.003 | 30 | 1042.5 | 15.465 | 15.468 |
| 29 | Xe IX $4d^{10}\ ^1S_0-4d^9 5p\ ^1P_1$ | 16.5323±0.003 | 30 | 1121.0 | 16.542 | 16.532 |
| 30 | Ar X $2s^2 2p^5\ ^2P^o_{3/2}-2s2p^6\ ^2S_{1/2}$ | 16.5530±0.02 | 38 | 1122.7 | 16.566 | 16.553 |
| 31 | Ne V $2p^2\ ^3P_2-2p3s\ ^3P^o_2$ | 16.7670±0.01* | 35 | 1138.3 | 16.784 | 16.767 |
| 32 | Ne IV $2p^3\ ^4S^o_{3/2}-2p^2(^3P)3d\ ^4P_{5/2}$ | 17.2620±0.01* | 35 | 1172.4 | 17.265 | 17.262 |
| 33 | Xe XXVI $4s\ ^2S_{1/2}-4p\ ^2P_{3/2}$ | 17.3938±0.005 | 39 | 1182.3 | 17.405 | 17.394 |
| 34 | Ne IV $2p^3\ ^2D^o_{5/2}-2p^2(^1D)3d\ ^2F_{7/2}$ | 17.7160±0.01* | 35 | 1204.7 | 17.725 | 17.716 |
| 35 | Ar XI $2s^2 2p^4\ ^3P_2-2s2p^5\ ^3P^o_2$ | 18.8820±0.02 | 38 | 1285.1 | 18.893 | 18.882 |

*wavelength accuracy estimated from the measurement technique, described in Ref. 35

**Figure Captions**

**Figure 1**. Photograph from EUV mirror (top) to CCD and schematic drawing (top view) of the EUV spectrometer for the EBIT facility at NIST. All linear dimensions are given in cm.

**Figure 2.** Detection efficiency calculated from SHADOW using a 300 μm x 3 cm EBIT source and a 300 μm slit, with and without Zr-filter (detector efficiency included as a multiplicative factor ≈0.42). As implemented, SHADOW overestimated the absolute efficiency of the grating, see main text for further discussion.

**Figure 3.** The solid dots are the positions of uniformly spaced monochromatic lines calculated using SHADOW. The solid line represents the fit function (Eq. 1) used to establish the relation between the linear and the wavelength position.

**Figure 4.** Schematic diagram showing the position $x_k$ of a known wavelength $\lambda_k$ in the grating focal plane (x-axis). The distance $x_0$ of the first CCD pixel from the origin is depicted.

**Figure 5.** EUV spectrum of Xe ions radiating from the EBIT operating at 8 keV with 147 mA of electron beam current, and entrance slits at 50 μm for 4 minutes. Individual numbered features are identified in Table II.

**Figure 6.** EUV spectrum of Ar ions radiating from the EBIT operating at 8 keV with 147 mA of electron beam current, and entrance slits at 50 μm for 12 minutes. Individual numbered features are identified in Table II.

**Figure 7.** EUV spectrum of Ne ions radiating from the EBIT operating at 8 keV with 147 mA of electron beam current, and entrance slits at 50 μm for 12 minutes. Individual numbered features are identified in Table II.

**Figure 8**. Comparison of two wavelength calibration procedures $\lambda_{syn}(p)$ and $\lambda_{std}(p)$ using the reported values $\lambda_{rep}$ listed in Table II. The shaded region represents the spectral width of a pixel (+/- _ pixel). Uncertainties shown are due only to the uncertainty in the $\lambda_{rep}$.

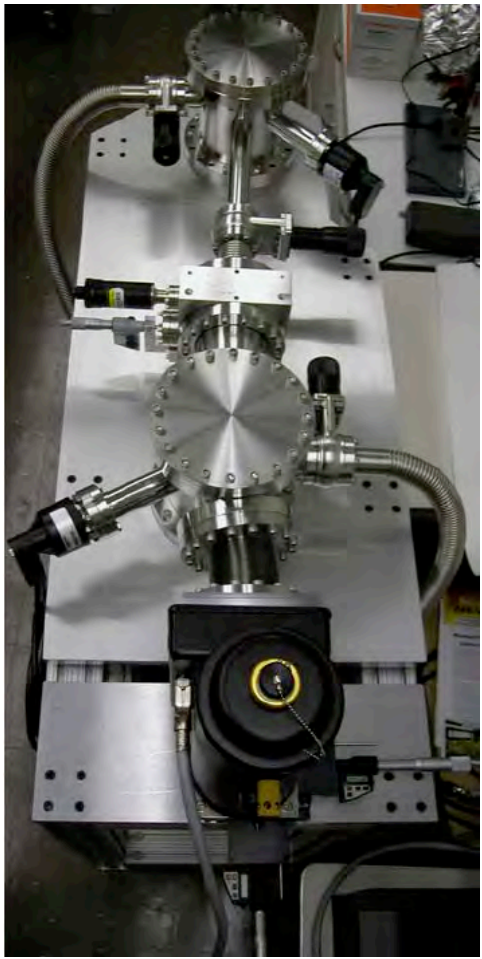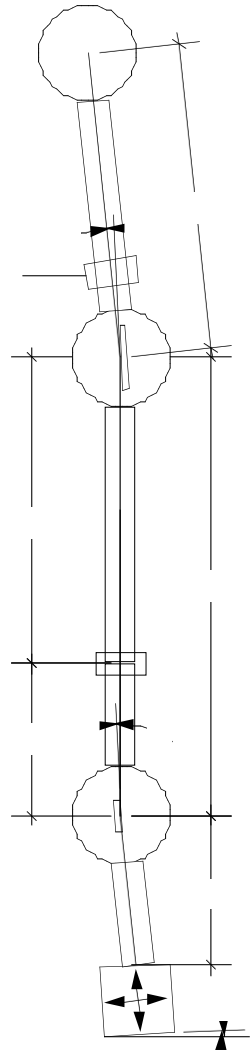

**Figure 1**.

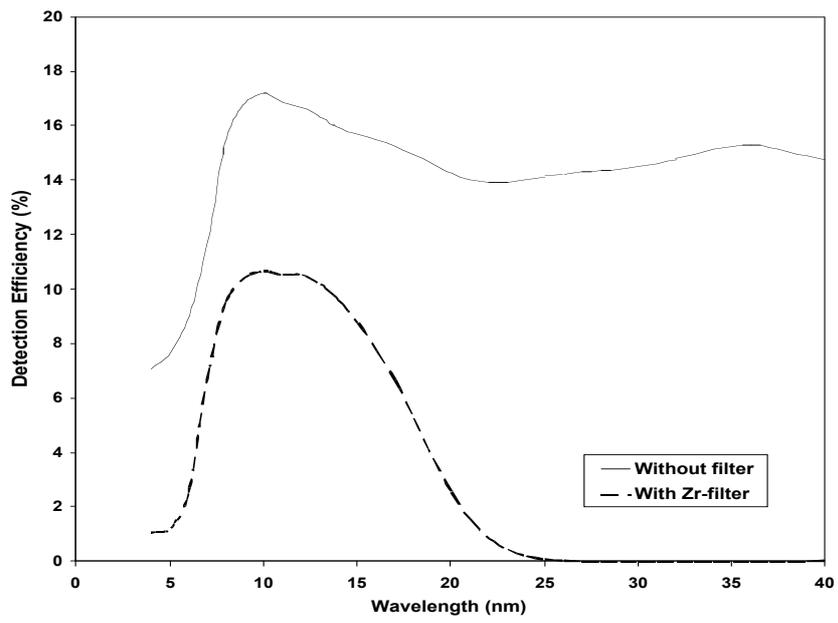

**Figure 2.**

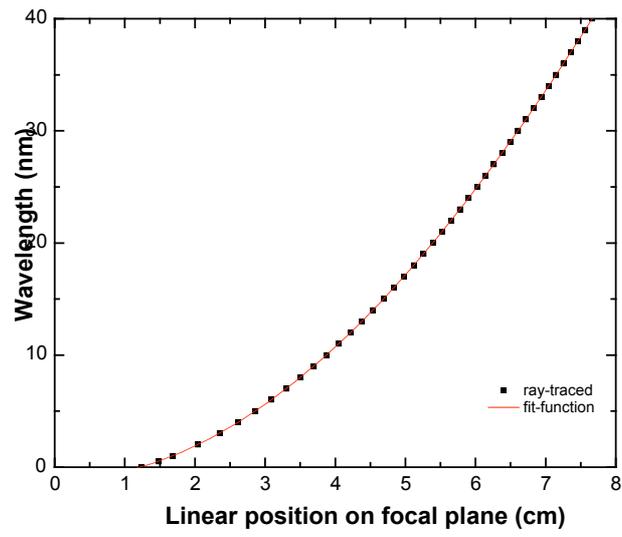

**Figure 3.**

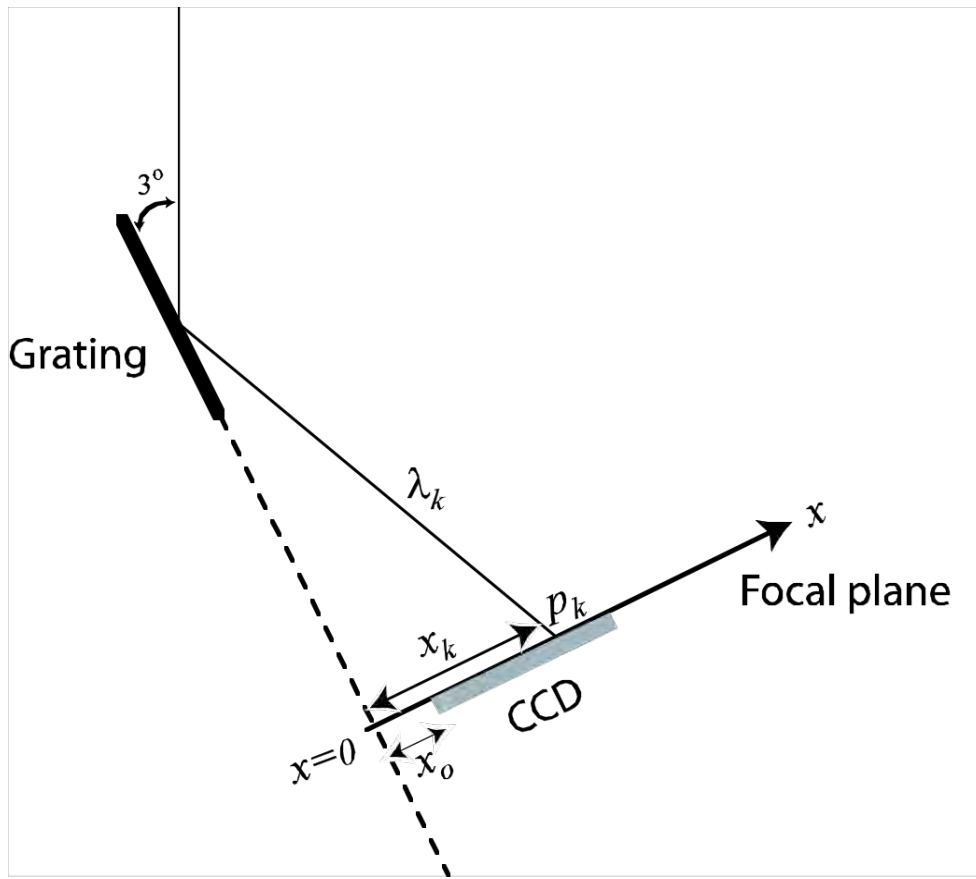

**Figure 4.**

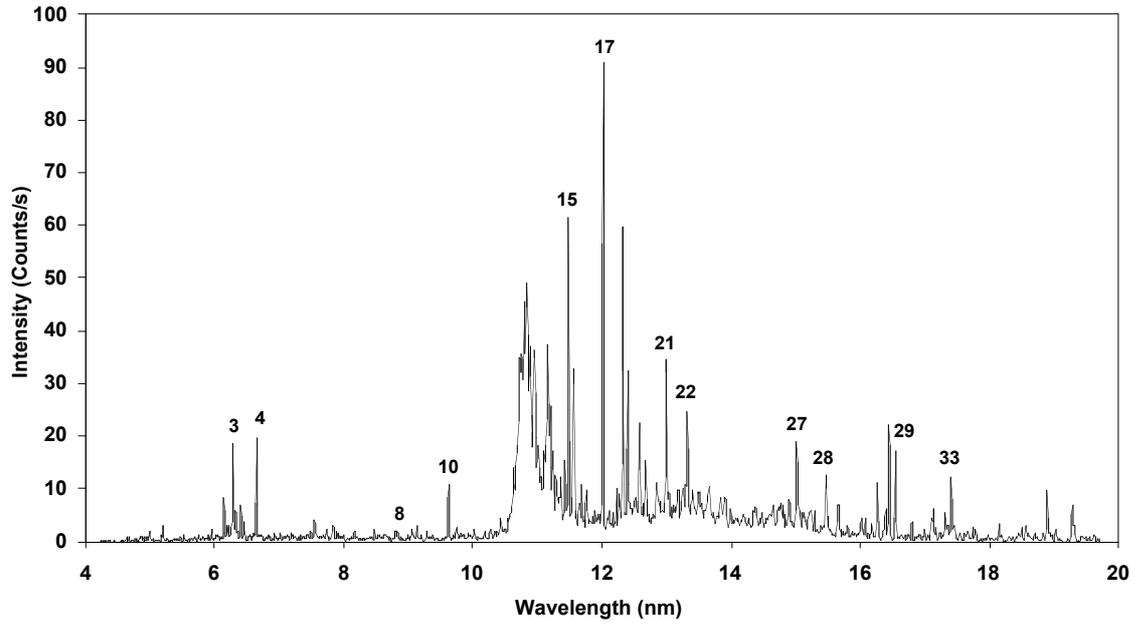

Xe @ 8 KeV, 137.1 mA, Slit = 50mm, t = 4 min

**Figure 5.**

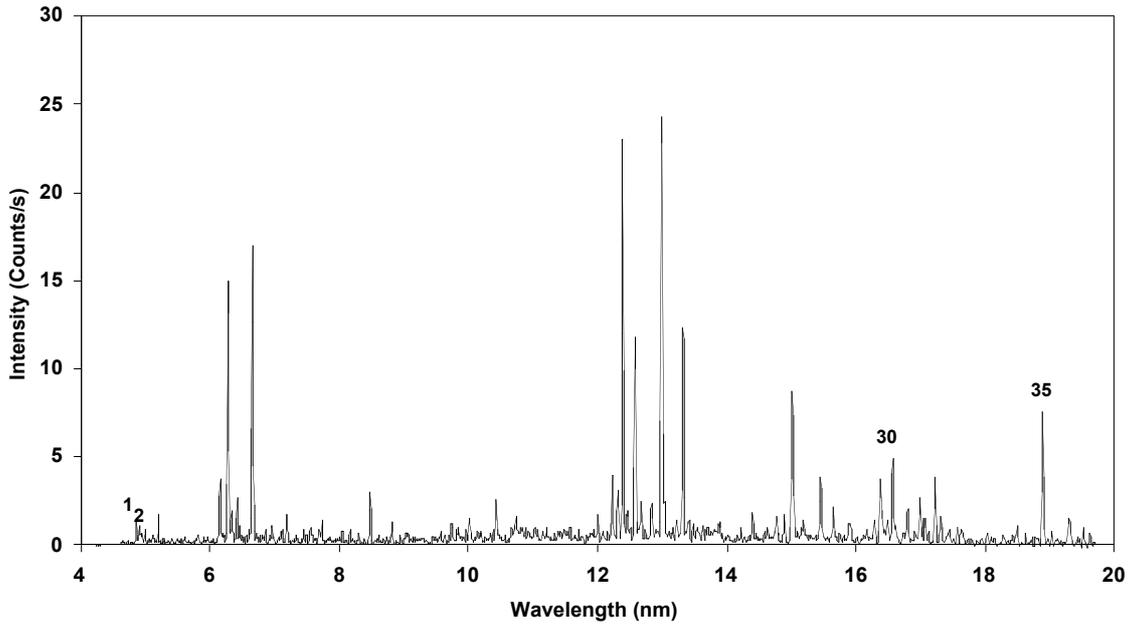

**Figure 6.**

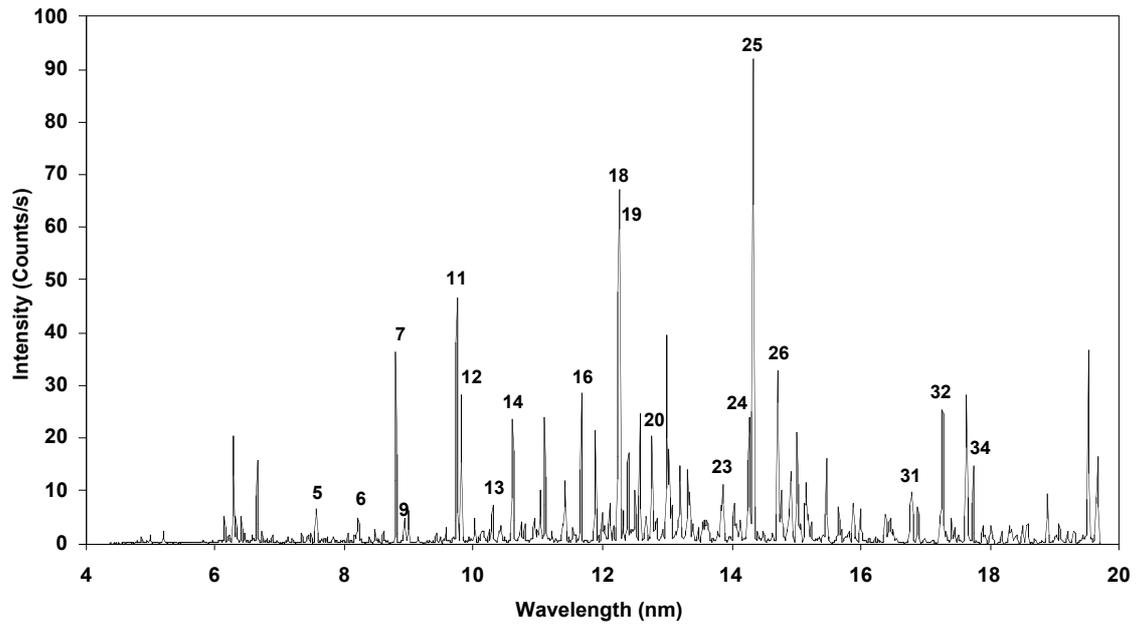

**Ne @ 8 KeV, 147 mA, Slit = 50mm, t = 12 min**

**Figure 7.**

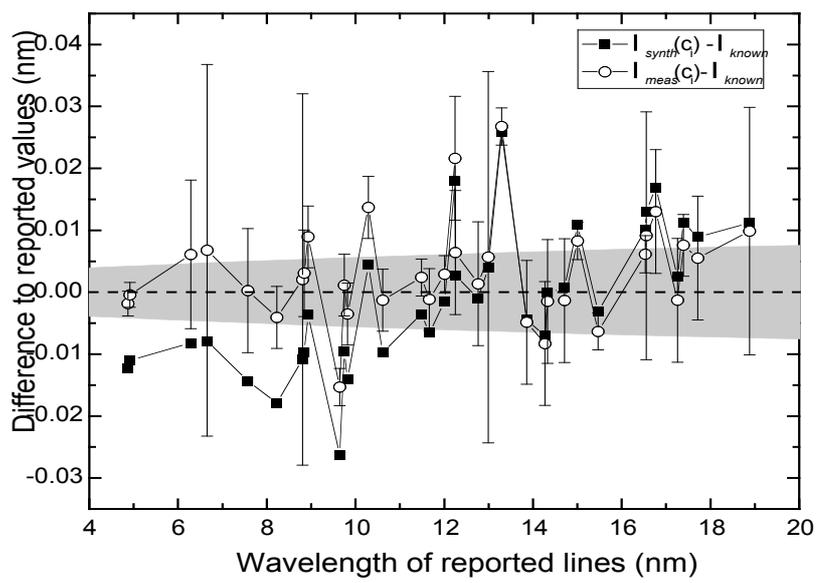

**Figure 8.**